\documentclass{svproc}

\usepackage{url}

\input{userDefs/myPreamble}
\input{userDefs/myCommands}

\begin{document}

\selectlanguage{british}

\mainmatter

\title{Towards indirect assessment of surface anomalies on wind turbine rotor blades}
\titlerunning{Towards indirect assessment of surface anomalies on rotor blades}

\author{Daniel~Feldmann\inst{1}\and
Felix~Oehme\inst{2}\and
Lennart~von~Germersheim\inst{1}\and
Rub{\'e}n~L{\'o}pez~Parras\inst{1}\and
Andreas~Fischer\inst{2}\and
Marc~Avila\inst{1}}
\authorrunning{Daniel~Feldmann \etal}
\tocauthor{Daniel~Feldmann, Felix~Oehme, Lennart~von~Germersheim,
Rub\'en~L\'opez~Parras, Andreas~Fischer, and Marc~Avila}

\institute{Universität Bremen,
Center of Applied Space Technology and Microgravity,\\
Am Fallturm 2, 28359 Bremen, Germany,\\
\email{daniel.feldmann@zarm.uni-bremen.de}
\and
Universität Bremen,
Institute for Metrology, Automation and Quality Science,\\
Linzer Straße 13, 28359 Bremen, Germany.}

\maketitle

\begin{abstract}
We present results from novel field, lab and computer studies, that pave the way
towards non-invasive classification of localised surface defects on running wind
turbine rotors using infrared ther\-mo\-gra\-phy (IRT). In particular, we first
parametrise the problem from a fluid dynamical point of view using the
roughness Reynolds number (\ReRough) and demonstrate how the parameter regime
relevant for modern wind turbines translate to parameter values that are
currently feasible in typical wind tunnel and computer experiments. Second, we
discuss preparatory wind tunnel and field measurements, that demonstrate a
promising degree of sensitivity of the recorded IRT data \wrt the key control
parameter (\ReRough), which is a minimum requirement for the proposed
classification technique to work. Third, we introduce and validate a local
domain ansatz for future computer experiments, that enables well-resolved
Navier--Stokes simulations for the target parameter regime at reasonable
computational costs.
\end{abstract}

\section{Background and motivation}

With \SI{132}{\tera\watt\hour} electricity fed into the grid in 2020, wind power
has become Germany’s leading source of energy (\SI{27}{\percent}) outperforming
any fossil source~\citep{Strom-Report2021}. However, cost reduction is
indispensable to further strengthen the competitiveness of sustainable energy
production beyond initial stages of massive subsidisation and political control.
Rotor blade inspection of ageing wind turbines, for example, is elaborate and
time-consuming and thus offers high potential to further reduce energy
production costs through more efficient maintenance procedures~\citep{Li2015a,
GarciaMarquez2020}.

\begin{figure}[htb]
\centering
\includegraphics[width=1.0\textwidth]{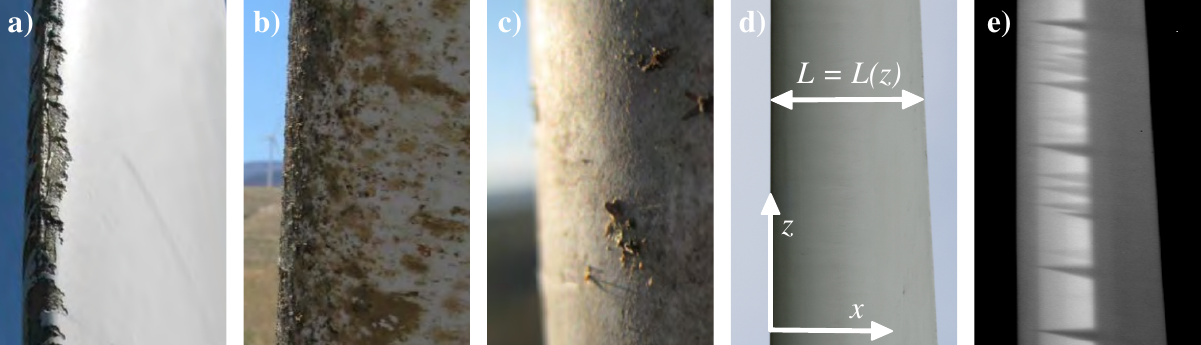}
\caption{Problem description. a--c): Wind turbine rotor blades with typical
surface defects as they occur during operation~\citep{Ehrmann2017}. a): Heavy
erosion of the blade's leading edge due to impact (\eg, sand, rain, hail). b,
c): Distributed and localised contamination with insect debris. d): Operating
wind turbine in our preliminary field measurements in Thedinghausen. The blade
suffers from naturally occurring surface anomalies, that are not visible from
the ground ($>\SI{100}{\meter}$). The streamwise direction is approximately in
$x$-direction and the blade's chord length ($L(z)\approx\SI{5}{\meter}$) varies
from root to tip. e): Thermographic image of the same blade (as in d), showing
regions with higher (light) and lower (dark) surface temperature (black is blue
sky compensation). Low temperature wedges indicate locally turbulent flow in the
downstream wake of small localised surface anomalies in contrast to the
otherwise laminar flow (high temperature).}
\label{fig:problem}
\end{figure}

\paragraph*{Health monitoring} of wind turbine blades usually employs visual
inspection (from the ground or using drones and climbers), in order to assess
damage and contamination level of the rotor blade's surface
(Fig.~\ref{fig:problem}a--c).
These methods require a shutdown of the turbine and/or suffer from poor spatial
resolution \citep{Kim2013, Shafiee2021}. Common alternatives (\eg strain-gauges,
optical and acoustical sensors) require elaborate instrumentation and suffer
from costly recertification~\citep{Li2015a}. Additionally, some of
these techniques heavily impair the aerodynamic performance of the rotor blades.
More advanced techniques (\eg ultrasonic, radiographic and electromagnetic
testing) are sometimes used in well-controlled lab environments, but are
essentially impossible to transfer to running wind turbines \mod{in} the field.

\paragraph*{Infrared thermography (IRT)} is an established diagnostic tool for
indirect flow visualisation in aerofoil boundary layers \citep{Gartenberg1992},
that overcomes the aforementioned drawbacks. It is based on the fact, that the
convective heat flux between the fluid and the surface strongly depends on the
local flow state (\eg laminar or turbulent). For a given temperature difference
between the incoming fluid and the aerofoil, flow state dependent spatial
temperature gradients develop on the surface. This gradients can be measured by
detecting the temperature dependent radiation from the surface with an infrared
camera. For example, IRT has been used in well-controlled wind tunnel
environments to detect laminar-turbulent transition and flow
separation~\citep{Montelpare2004, Dollinger2018a, Dollinger2019}. In these type
of studies, the aerofoils are actively heated with
\mod{$\orderof{\num{e3}}\si{\watt\per\meter\squared}$}, to increase the
convective heat flux density and thereby improving the thermal contrast (TC)
between regions of different flow states. In contrast, applying IRT to operating
wind turbines in the field is much more challenging because of several
reasons~\citep{Dollinger2018a}. First, active rotor blade heating is undesired.
This heavily limits the maximum achievable TC, which now depends on local
weather conditions and absorbed solar irradiation (\ie
\mod{$\orderof{\num{e2}}\si{\watt\per\meter\squared}$} during clear sky).
Second, measuring from remote ground locations ($> \SI{100}{\meter}$) causes
large transmission losses and low contrast to noise ratios (CNR). Third, the
camera exposure time must not exceed $\orderof{\num{e-4}}\si{\second}$, in order
to prevent blur due to rotor blade motion. Despite these constraints, IRT has
been recently used to successfully detect premature laminar-turbulent flow
transition due to surface anomalies in operating wind turbines
\citep{Traphan2018, Reichstein2019, Gleichauf2020, Parrey2021}. These
achievements open new avenues to monitor damage and contamination level of wind
turbine blades (Fig.~\ref{fig:problem}a--c) via IRT of the blade's surface
(Fig.~\ref{fig:problem}e) without shutting the turbine down for inspection nor
modifying the rotor with additional sensors. For the application, however, it is
not sufficient to simply detect the presence of an anomaly via IRT. Instead, it
would be necessary to additionally infer exact position, size and shape of
geometrical defects in order to enable remote classification of the surface
anomaly; \eg distinguish between erosion and deposition
(Fig.~\ref{fig:problem}a,b) or distinguish between different levels of
contamination (Fig.~\ref{fig:problem}b,c). This poses many methodological
challenges and it is currently unclear, to what extent such a classification can
be inferred faithfully from IRT images and whether physically informed modelling
approaches might improve this endeavour.

\paragraph*{Here we present} results from our preliminary studies paving the way
towards indirect assessment of surface anomalies on running wind turbine rotor
blades. In \S\ref{sec:state}, we briefly summarise the state of the art in early
laminar-turbulent transition in flat plate (Blasius) boundary layers and
establish terminology necessary for the rest of the paper. In \S\ref{sec:param},
we introduce a parametrisation of our problem in terms of the roughness Reynolds
number (\ReRough), which is commonly used to characterise roughness induced
transition in Blasius boundary layers. In \S\ref{sec:preIRT}, we discuss new
results from our wind tunnel and field measurements, that demonstrate a
promising degree of sensitivity of the recorded IRT data with respect to
\ReRough. In \S\ref{sec:preDNS}, we introduce and validate a local domain ansatz
for future computer simulations, that enables well-resolved 3d DNS for the
target parameter regime at reasonable computational costs. In \S\ref{sec:per},
we briefly propose future directions.

\section{Turbulence transition in Blasius' boundary layer flow}
\label{sec:state}

Probably the best understood example of boundary layer (BL) transition is a flat
plate BL flow. In this canonical \mod{benchmark} system, the streamwise velocity
profile, $u_x=u_x(x,y)$, and thus the thickness of the BL, $\delta_{99}(x)$, are
both very well described by the self-similar Blasius solution
\citep[\eg][]{Puckert2018}. The Blasius profile only depends on the wall-normal
coordinate ($y$) and can be re-scaled for arbitrary streamwise locations ($x>0$)
just by fixing the Reynolds number, $\ReChord=\sfrac{x\,u_\infty}{\nu_\infty}$,
where $u_\infty$ and $\nu_\infty$ are the velocity and the viscosity of the
fluid in the far field. \par
The Blasius BL becomes linearly unstable at $\ReChord=\orderof{10^5}$. In
experiments, however, this instability is bypassed and laminar-turbulent
transition can occur at lower \ReChord due to finite-amplitude perturbations,
which are usually present in the free-stream \citep{Brandt2004}. Especially
relevant to our present work is the transition due to single (isolated)
geometric perturbation elements on the plates surface, which perturb the
velocity profile locally. Following a number of studies chiefly employing flow
visualisations \citep{Klanfer1953, Smith1959}, Klebanoff \etal
\citep{Klebanoff1992} provided an in depth, quantitative analysis of the
transition to turbulence caused by a semisphere glued to a flat plate's surface.
They showed that the key parameter governing the transition is the roughness
Reynolds number, $\ReRough=\sfrac{u_k\,k}{\nu_\infty}$, where $k$ is the height
of the roughness element (\eg radius of the semisphere) and
$u_k=u_x(x=x_k,y=k)$ is the undisturbed streamwise velocity component at that
height ($y=k$) and streamwise location ($x=x_k$). They placed semispheres with
different $k$ at several $x_k$ and found a critical Reynolds number of
$\ReRough\approx\num{325}$, that is independent of $x_k$. They also studied a
cylinder of unit aspect ratio ($\Gamma_k=\sfrac{k}{d}=1$, where $d$ is the
cylinders diameter) and found a critical Reynolds number of
$\ReRough\approx\num{450}$, which was lower than in most previous studies
\citep{Smith1959}. Since then, many experimental
\citep{Ye2016, Puckert2018, Ye2018} and numerical
\citep{Masad1994, Fransson2005, Loiseau2014, DeGrazia2018, Casacuberta2020}
studies have been carried out, and it has been repeatedly highlighted that
\ReRough is the main parameter governing transition, albeit the geometry of the
element (including $\Gamma_k$) plays an important role as well. For the specific
case of cylinders, Puckert \& Rist \citep{Puckert2018} demonstrated a changeover
from purely convective to global instability for transcritical \ReRough and an
additional competition between varicose and sinuous instabilities, when the
element is especially thin (\ie large $\Gamma_k$). This indicates that the
perturbation element solely acts as an amplifier in the subcritical regime
(leading to symmetrical wake structures) and, additionally, as a wavemaker in
the supercritical regime (leading to wake vortex shedding). We expect, that
already these subtle differences, for example, will be distinguishable in the
instantaneous wall-shear stress and thus in the plate's surface temperature
distribution as well, assuming a unit Prandtl number and Reynolds' analogy to
hold. \par
For wind turbines, however, we expect the curvature of the blade's surface to
play an important role for this specific transition scenario. First, because the
curved geometry results in streamwise pressure gradients and velocity profiles
that can deviate notably from the Blasius solution. Second, because the velocity
profile is not self-similar, implying that a particular \ReRough can result from
different combinations of $(x_k,k)$, as discussed in \S\ref{sec:param}. It is
currently unclear to what extent these conditions change the underlying
mechanisms (compared to Blasius' scenario) and whether these changes might help
or hinder our endeavour.

\section{Relevant parameter space (\ReChord-\ReRough)}
\label{sec:param}

\begin{figure}[tb]
\centering
\includegraphics[width=1.00\textwidth]{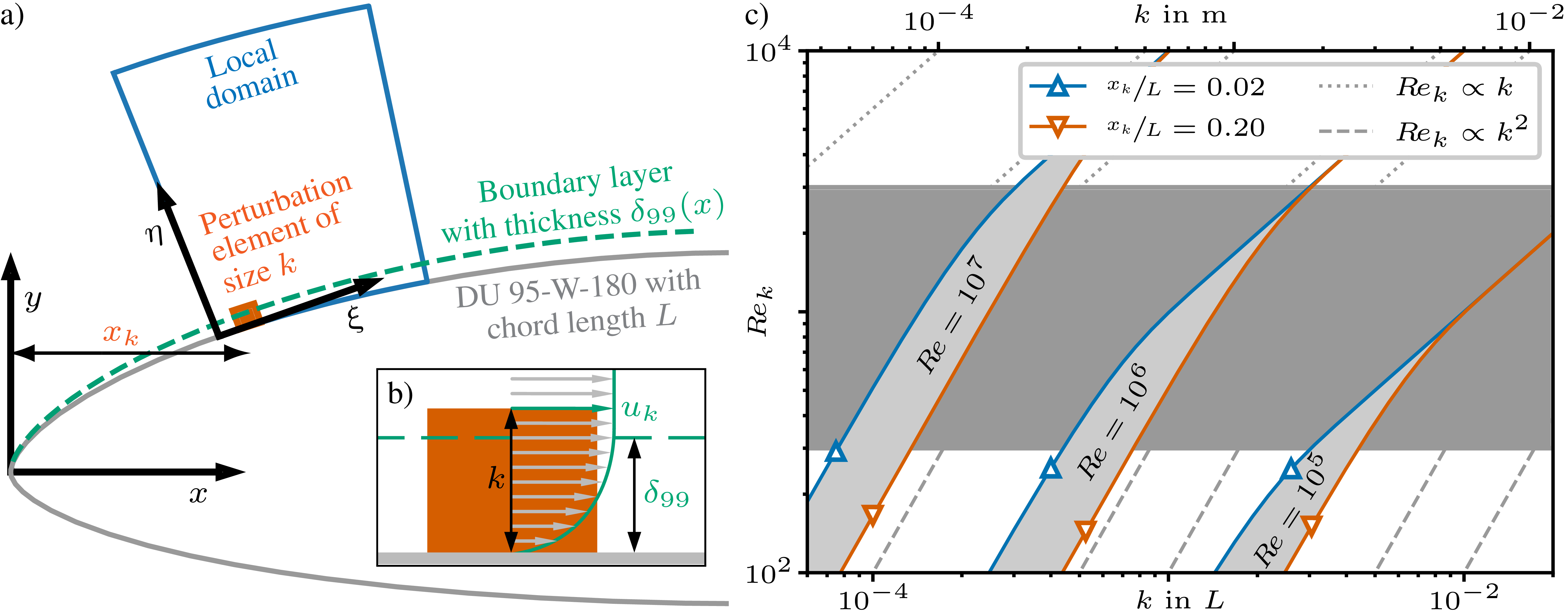}
\caption{Parametrisation of the problem. a): Sketch of a typical wind turbine
rotor blade profile (DU 95-W-180 \citep{Timmer2003}) with its leading edge at
$(x=0,y=0)$ in the global coordinate system. A laminar boundary layer (broken
line) is growing from its leading edge and a generic perturbation element is
located at $\sfrac{x_k}{L}\approx\sfrac{1}{10}$. A local computational domain is
placed around the perturbation element. \mod{b): Zoom of a.} c): Relevant
parameter space in terms of element size ($k$), position ($x_k$), chord
(\ReChord) and roughness (\ReRough) Reynolds number, based on the crude
assumption of a Blasius \mod{BL to estimate the local BL thickness,
$\delta_{99}=\delta_{99}(x_k)$, and the relevant velocity scale,
$u_k=u_k(x_k, k)$}.}
\label{fig:param}
\end{figure}

Modern wind turbines in the multi-megawatt class operate at Reynolds numbers up
to $\ReChord=\num{e7}$, where $x$ is taken to be $L$ (\ie the local chord length
of the rotor blade), and $u_\infty$ is taken to be the relative wind speed at a
particular radial location ($z$) of the rotating blade (Fig.~\ref{fig:problem}e)
\citep{Ge2014}. Modern wind turbines also employ rather thick rotor blade
profiles (Fig.~\ref{fig:param}a)~\citep{Timmer2003}, so that even for large
angle of attack ($\alpha$), the most prone location for surface anomalies due to
foreign impact remains the vicinity of the blade's leading edge
($\sfrac{x}{L}\lesssim\num{0.2}$). In this region, the BL is typically laminar,
since natural transition in wind turbine rotor blades usually occurs later
($\sfrac{x}{L}\gtrsim\num{0.3}$). Now, we here consider generic geometric
perturbation elements with a characteristic size ranging from
$k=\orderof{10^{-5}}\si{\meter}$ (reflecting \eg small sand grain) up to
$k=\orderof{10^{-2}}\si{\meter}$ (reflecting \eg insect debris) in order to
model highly localised deposition on operating wind turbine rotor blades
(Fig.~\ref{fig:problem}c). Note, that more severe (distributed) contamination
(Fig.~\ref{fig:problem}b) and surface ablations (Fig.~\ref{fig:problem}a) will
be considered elsewhere. Furthermore, we begin by making the crude assumption,
that the laminar BL in this region can be approximated by Blasius' solution.
This allows us, for the first time, to roughly quantify the relevant range of
\ReRough for our problem by estimating the a priori unknown $u_k$ from the given
parameter values (\ReChord, $x_k$, $k$) discussed above. \par
The results are summarised in figure~\ref{fig:param}c.
The dark grey region denotes the range of \ReRough realised in our preliminary
wind tunnel campaign (at $\ReChord=\num{e5}$, see \S\ref{sec:preIRT}), covering
one order of magnitude from the sub-critical ($\ReRough\approx\num{300}$) up to
the fully turbulent ($\ReRough\approx\num{3000}$) regime.
Figure~\ref{fig:param}c also indicates, how these values translate to other
situations; \eg, other combinations of $(x_k,k)$ or higher \ReChord. In this
context, it is important to note, that for $k\ll\delta_{99}$ (\ie element fully
embedded in the linear part of the Blasius BL), \ReRough scales quadratic with
$k$ for a given location ($x_k$), whereas for $k\gg\delta_{99}$ (\ie element
penetrates the flow outside the Blasius BL, as in fig.~\ref{fig:param}b),
\ReRough scales only linear with $k$ and is thus basically independent of $x_k$.
Where exactly this transition from the quadratic to the linear scaling takes
place, depends on \ReChord. In any case, relevant locations for generic
perturbation elements would be $\num{0.02}<\sfrac{x_k}{L}<\num{0.2}$, according
to the discussion above. For a thick rotor blade profile (in contrast to the
crude Blasius assumption here), the scaling in this two regimes will not be
exactly quadratic and linear, due to curvature effects, but we expect the
general behaviour to be qualitatively similar.

\section{Infrared thermography (IRT)}
\label{sec:preIRT}

We have performed preparatory IRT measurements on a running
\SI{1.5}{\mega\watt}-wind turbine in Thedinghausen, Germany, and on a
down-scaled wind tunnel model using the \windguard facilities near Bremerhaven.
Our results show, that it is feasible to distinguish between different localised
perturbation elements in terms of their size ($k$) and location ($x_k$) by means
of an IRT image of the blades' surface temperature.\par
Already our field measurement at $\ReChord=\orderof{\num{e6}}$ demonstrates a
very promising degree of sensitivity of the IRT method \wrt different
perturbations in the vicinity of the blade's leading edge
(Fig.~\ref{fig:problem}e). Note, that here natural perturbations on the rotor
blade's surface (as they occur during operation) generate a variety of
distinguishable thermal wake patterns in the IRT image. In particular, the
different shades of dark grey wedges indicate regions of different temperatures
at different locations on the blade's surface that are presumably generated by
surface anomalies of different size and location.\par
\begin{figure}
\includegraphics[width=1.0\textwidth]{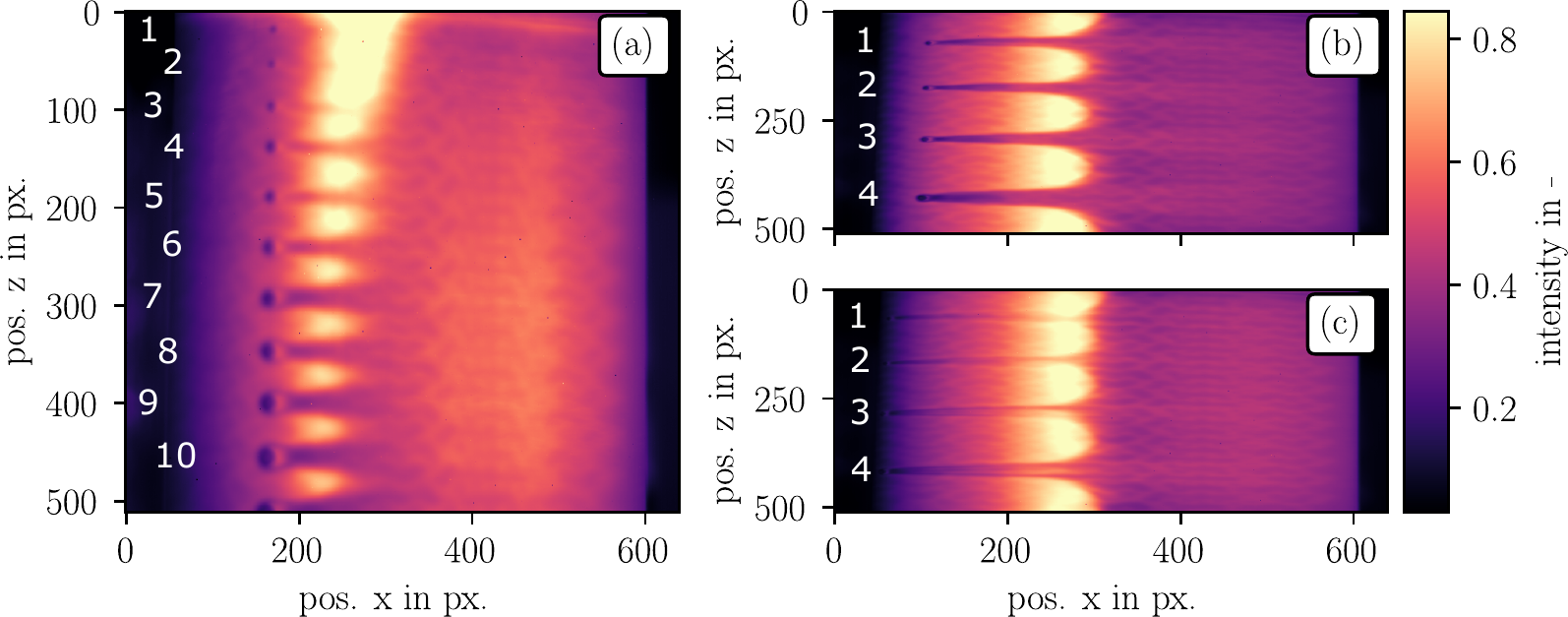}
\caption{Sensitivity analysis in the wind tunnel. Shown are IRT images of the
blade's surface for fixed $(\ReChord=\num{e5}, \alpha=\SI{2}{\degree})$ and
varying $(\ReRough,x_k)$, where higher/lower intensity reflects a warmer/cooler
surface. The blade geometry is a DU 97-W-250 profile \citep{Timmer2003} with a
constant chord length of $L=\SI{30}{\centi\meter}$. a): Sensitivity of the
thermogram for fixed $(\sfrac{x_k}{L}=\num{0.2})$ and varying
$\ReRough\in[\num{270},\num{800}]$ in the transcritical regime, where we expect
the onset of unsteady flow. The cylindrical perturbation elements are numbered
from one to ten with increasing \ReRough. b): Same as in a, but for
$(\sfrac{x_k}{L}=\num{0.1})$. The four elements are numbered with increasing
$\ReRough\in[\num{800},\num{3100}]$. c): Same as in b, but for
$(\sfrac{x_k}{L}=\num{0.03})$.}
\label{fig:preIRT}
\end{figure}

Our wind tunnel measurements at $\ReChord=\orderof{\num{e5}}$ are a first
successful effort to better quantify the underlying relationships between
surface anomalies (cause) and their footprint in the surface temperature
(effect). Here, we equipped the suction side of a typical rotor blade profile
(DU~97-W-250~\citep{Timmer2003}, $L=\SI{30}{\centi\meter}$,
$\alpha=\SI{2}{\degree}$) with ten cylindrical perturbation elements of varying
size ($k\in[\num{0.5},\num{1.6}]\si{\milli\meter}$, constant
$\Gamma_k=\num{0.3}$) and repeated the IRT measurements for three different
element locations ($\sfrac{x_k}{L}\in\{\num{0.03},\num{0.1},\num{0.2}\}$). Our
results are summarised in figure~\ref{fig:preIRT}.
Although \ReRough is the only control parameter in the flat plate transition
scenario, it is apparent from fig.~\ref{fig:param}, that on an aerofoil, one
particular \ReRough can be realised with different combinations of $(x_k,k)$ for
given free-stream conditions (\ReChord). Thus, we first fix
$\sfrac{x_k}{L}=\num{0.2}$ to isolate the effect of varying $k$
(Fig.~\ref{fig:preIRT}a). In this particular scenario, IRT is remarkably
sensitive with respect to a varying $k$, which here corresponds to
$\ReRough\in[\num{270},\num{800}]$. From element \#3 on (\ie\ReRough>\num{500}),
a thermal signature is clearly visible in the downstream wake
(Fig.~\ref{fig:preIRT}a). This indicates that IRT is indeed capable of detecting
weak transitional wakes at low trans-critical \ReRough. If we now translate this
first rough estimate of a minimum threshold for detectability
($\ReRough\approx\num{500}$) to a real-word wind turbine scenario by making use
of fig.~\ref{fig:param}b, we can conclude that our proposed measurement approach
is suitable for detecting rotor blade defects as small as roughly
\SI{0.1}{\milli\meter}, with the exact threshold depending on both,
$\sfrac{x_k}{L}$ and \ReChord. Additionally, the size of the wake pattern as
well as its CNR \wrt the undisturbed (\ie warmer) surrounding increase with
\ReRough (Fig.~\ref{fig:preDNS}a), thereby demonstrating a continuous
sensitivity of our measurement approach.\par
Comparing figs.~\ref{fig:preIRT}a, \mod{b, and c} additionally demonstrates that
the IRT method is clearly capable of distinguishing different $x_k$ in the
entire range of relevant distances from the blade's leading \mod{edge}
($\sfrac{x_k}{L}\in[\num{0.03},\num{0.2}]$). However, for elements at
$\sfrac{x_k}{L}=\num{0.03}$, the wake pattern in the thermogram is less distinct
when compared to the wake patterns originating from elements placed farther
downstream (\eg $\sfrac{x_k}{L}=0.1$). In particular, the CNR of the thermal
wake patterns increases twofold, when the perturbation elements are moved
farther downstream (\ie from $\sfrac{x_k}{L}=0.03$ to $\sfrac{x_k}{L}=0.1$). We
conclude, that the position of a surface defect directly affects the
detectability of its thermal wake pattern. Note, that for the four elements
shown in figs.~\ref{fig:preIRT}b and c, $\ReRough\in[\num{800},\num{3100}]$,
which is considerably higher than the roughness Reynolds numbers shown in
figure~\ref{fig:preIRT}a.

\section{Direct numerical simulations (DNS)}
\label{sec:preDNS}

We have performed preparatory DNS of the incompressible Navier--Stokes equations
using the spectral-element framework \nektar, which is purposely designed for
high-fidelity flow simulations on unstructured grids in moderately complex
domains. In order to enable well-resolved 3d DNS of laminar-turbulent transition
in a rotor blade boundary layer with geometric perturbation elements at
reasonable computational costs, we introduce and validate a \emph{local domain
ansatz} in massively reduced computational domains. We stress, that otherwise an
extensive variation study of the relevant control parameters would be
prohibitively expansive for the intended range of parameter values
($\ReChord>\num{e5}$, $\ReRough\gg\num{300}$). \par
The general idea of our local ansatz (Fig.~\ref{fig:param}a) goes as follows.
First, we perform a 2d simulation of a typical wind turbine rotor blade
(DU~95-W-180~\citep{Timmer2003}) in a large (global) computational domain
without any perturbation element. Note, that we here use a sufficiently large
domain (C-type, dimensions $\num{15}L\times\num{12}L$), in order to generate
tailored inflow and pressure boundary conditions (BC) for much smaller local
domains, that are located around the desired $x_k$. \mod{Recall, that the flow
in the region of interest (\ie $\sfrac{x}{L}<\num{0.2}$) is always laminar,
justifying our 2d approach}. In a second step, we perform well-resolved 3d DNS
in a local domain capturing only a small part of the curved rotor blade surface.
In the future, this will allow us to \mod{include} different isolated
perturbation elements and, for the first time, investigate the
transitional/turbulent flow in their downstream wake and the resulting local
heat transfer between the fluid and the curved rotor blade surface in
unprecedented detail. \par
\begin{figure}[bt]
\centering
\includegraphics[width=1.0\textwidth]{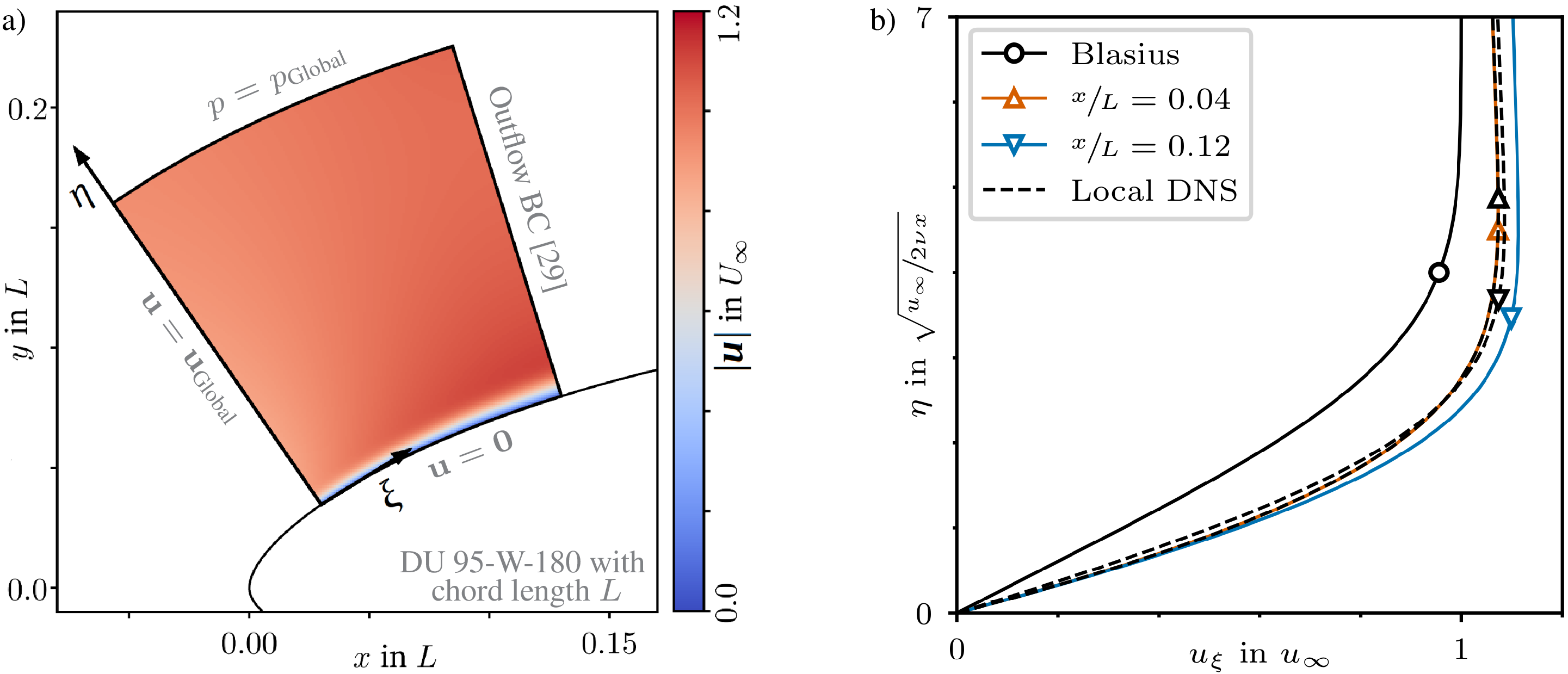}
\caption{Local domain ansatz. a): Velocity field
($\lvert\vec{u}\rvert=\sqrt{u_x^2+u_y^2}$)
generated in a local 2d simulation ($\ReChord=\num{e4}$, domain at
$\num{0.03}\le\sfrac{x}{L}\le\num{0.13}$) of a clean rotor blade
(DU~95-W-180~\citep{Timmer2003}, $\alpha=\SI{2}{\degree}$, w/o perturbation)
\mod{using Dirichlet BC for $\vec{u}$ and $p$ as indicated and a robust outflow
BC \cite{Dong2014} at the right}. b): Comparison of the velocity profile,
$u_\xi(\eta)$, in the boundary layer at two different streamwise locations,
$\sfrac{x}{L}$, as predicted in local (broken lines) and global (full lines)
DNS. The Blasius solution is given as reference.}
\label{fig:preDNS}
\end{figure}

Here we demonstrate the general viability of our ansatz (Fig.~\ref{fig:preDNS}).
For example, fig.~\ref{fig:preDNS}b compares velocity profiles from a global and
a local DNS run (here both 2d, $\ReChord=\num{e4}$) and shows two things. First,
the velocity profiles in the rotor blade boundary layer are, as expected,
substantially different from the Blasius solution. Second, the results obtained
from local DNS runs agree very well with the global DNS results. In particular,
close to the inflow BC of the local domain ($\sfrac{x}{L}=\num{0.04}$, where
later perturbation elements will be placed), the velocity profiles are identical
(judged by the given line width of the plot). Close to the outflow BC of the
local domain ($\sfrac{x}{L}=\num{0.12}$), the BL profiles have slightly diverged
\mod{($||u_{\zeta,\text{local}}-u_{\zeta,\text{global}}||_\infty\le\num{e-2}$)},
but still show very goo{d qualitative agreement. Note, that these results
represent a worst-case scenario, since the local domain starts very close to the
blade's leading edge, where surface curvature has the strongest effect. For
local domains that are located farther downstream (\eg starting at
$\sfrac{x}{L}=\num{0.1}$ in order to capture perturbation elements farther away
from the leading edge), we find that the deviations between local and global DNS
results are even smaller (not shown). \mod{Note, that we have tested and
analysed many different combinations of BC for the local DNS ansatz. In all
cases, we have used Dirichlet and Neumann BC for $u$ and $p$, respecticaly, at
the inflow (left) and the blades surface (bottom), where the values for the
Dirichlet BC were extracted from the global run. The results presented in
fig.~\ref{fig:preDNS}, we obtained using pressure Dirichlet BC at the top and
the robust outflow BC of Dong \etal~\cite{Dong2014} at the rigth. Other
combinations of BC for the top/right boundaries yield very similar results
with locally slighty better agreement, depending on the region of interest (\ie
either inside or outside the BL).}

\section{Perspective}
\label{sec:per}

Our preliminary studies show two things. First, IRT is sufficiently sensitive to
distinguish between different size and location of highly localised perturbation
elements in typical wind turbine rotor blade boundary layers. Second, the
proposed local domain ansatz is feasible to approximate the boundary layer flow
on a typical rotor blade profile sufficiently well, allowing for well resolved
Navier--Stokes simulations of the underlying problem at reasonable
computational costs. \par
Future works should aim at generating a comprehensive surface-state-flow-pattern
catalogue and a deeper physical understanding of how surface irregularities in
the vicinity of the blade's leading edge (cause) affect the downstream surface
temperature distribution (effect). Here, it is particularly interesting to
clarify, how much information about the three-dimensional flow in the wake is
actually contained in its two-dimensional surface temperature projection, since
the inverse problem (effect to cause) is not bijective. To this end, extensive
wind tunnel and computer experiments could be combined in order to
systematically characterise the effect of different generic surface anomalies on
early transition and the resulting (turbulent) wake in a rotor blade boundary
layer and the resulting temperature footprint. Hopefully, this will lay the
foundation for the development of a data-trained physics-informed model to
reliably relate individual thermographic images to particular rotor blade
conditions based on image recognition and machine learning techniques.

\paragraph*{We gratefully acknowledge} resources and support provided by Nicolas
\mbox{Balaresque} (\windguard) and the \hlrn.

\bibliographystyle{spphys}
\bibliography{stab2022.bib}

\end{document}